\documentclass[conference]{IEEEtran}
\IEEEoverridecommandlockouts
\usepackage{cite}
\usepackage{amsmath,amssymb,amsfonts}
\usepackage{algorithmic}
\usepackage{graphicx}
\usepackage{textcomp}
\usepackage{xcolor}
\usepackage{multirow}
\usepackage[table]{xcolor}
\def\BibTeX{{\rm B\kern-.05em{\sc i\kern-.025em b}\kern-.08em
    T\kern-.1667em\lower.7ex\hbox{E}\kern-.125emX}}
\begin{document}




\title{Experimental Evaluation of Fuzzy-Integral and Classical controls for Power Management in a 24 GHz mmWave 5G Transceiver
}

\author{

    \IEEEauthorblockN{Karel Walter Gomez Orellana}
    \IEEEauthorblockA{\textit{Ingeniería Electrónica} \\
        \textit{Universidad Mayor de San Andrés}\\
        La Paz, Bolivia \\
        karelgomez124@ieee.org}
    \and

    \IEEEauthorblockN{Berthyn Rodrigo Tiñini Chuquimia}
    \IEEEauthorblockA{\textit{Ingeniería Electrónica} \\
        \textit{Universidad Mayor de San Andrés}\\
        La Paz, Bolivia \\
        berthyn.tinini@ieee.org}
    \and
    
    \IEEEauthorblockN{Juan Carlos Paredes Condori}
    \IEEEauthorblockA{\textit{Ingeniería Electrónica} \\
        \textit{Universidad Mayor de San Andrés}\\
        La Paz, Bolivia \\
        jcparedes7@umsa.bo}
    \and
    
    \IEEEauthorblockN{\hspace{3.5 cm}Rodrigo Apaza Huanca}
    \IEEEauthorblockA{\textit{\hspace{3.5 cm}Ingeniería Electrónica} \\
        \textit{\hspace{3.5 cm}Universidad Mayor de San Andrés}\\
        \hspace{3.5 cm}La Paz, Bolivia \\
        \hspace{3.5 cm}rapaza94@umsa.bo}
    \and
    
    \IEEEauthorblockN{Hugo Orlando Condori Quispe}
    \IEEEauthorblockA{\textit{Ingeniería en Sistemas} \\
        \textit{Electrónicos y de Telecomunicaciones}\\
        \textit{Universidad Privada Boliviana}\\
        \textit{Ideas RF}\\
        La Paz, Bolivia \\
        hugo.condori@fulbrightmail.org}
}

\maketitle

\begin{abstract}
The deployment of 5G millimeter-wave (mmWave) systems poses significant challenges in maintaining power amplifier linearity and efficiency under varying conditions, such as temperature-induced gain variations that degrade error vector magnitude (EVM). This paper presents a comparative study of three control strategies—PID, pure integral, and fuzzy-integral (FI)—for adaptive power management in a 24 GHz mmWave transceiver. The FI controller integrates fuzzy logic for handling nonlinearities with integral action for zero steady-state error. Experimental results show the FI controller outperforms others in settling time, stability, and EVM minimization.
\end{abstract}


\section{Introduction}
The millimeter-wave (mmWave) bands are pivotal for 5G New Radio (NR). However, mmWave transceivers face challenges in maintaining power amplifier (PA) linearity due to high peak-to-average power ratios (PAPR) in OFDM waveforms (e.g., with 64/256-QAM subcarriers). Nonlinear distortion degrades error vector magnitude (EVM), and causes spectral regrowth.

To mitigate the effects of high PAPR and nonlinearities, it is essential to actively control the PA output power so it operates in the linear region/back-off. Prior works have proposed integrated power detection architectures to enable such precise control in mmWave 5G transmitters \cite{b1}, and fuzzy-based control schemes have been shown effective for adaptive resource and power management in 5G networks \cite{b4}. By maintaining linear operation, the system ensures low EVM and controlled spectral purity even under varying input conditions. Beyond input-drive dynamics, ambient temperature fluctuations and the effective electrical length of interconnecting cables can introduce insertion-loss and gain drifts of several decibels along the RF path: (i) for long or thermally graded coaxial runs, measured S-parameters show that the transmission coefficient |S21| changes significantly under large temperature excursions, directly impacting end-to-end gain \cite{colangeli}; and (ii) for active stages, electro-thermal effects reduce PA gain with temperature—ranging from sub-dB dynamic drops over milliseconds to multi-dB shifts over wide temperature spans—which in turn degrades EVM if left uncompensated \cite{Taghikhani,Najjari}.

Conventional control strategies, such as proportional-integral-derivative (PID) controllers and automatic gain control (AGC) loops, have been widely employed for power management in RF transceivers, including mmWave systems \cite{b5,b6}. However, these classical methods often struggle with the inherent nonlinearities, uncertainties, and rapid variations in mmWave environments, leading to suboptimal performance in terms of settling time and stability. In particular, when gain drifts are driven by temperature transients and cable-run losses, maintaining linear-region operation and EVM targets demands thermal-aware compensation or adaptivity in the loop \cite{Taghikhani,Najjari}. 

This paper advances these efforts by introducing a fuzzy-integral (FI) controller, which combines fuzzy logic's robustness to nonlinearities and uncertainties with integral action for zero steady-state error, building upon prior work on fuzzy-based control in beyond 5G networks \cite{b2,b3}. The FI controller is designed and implemented on hardware, showing a superior dynamic performance.

\section{System Architecture}
\begin{figure}[htbp]
\centerline{\includegraphics[width=0.48\textwidth]{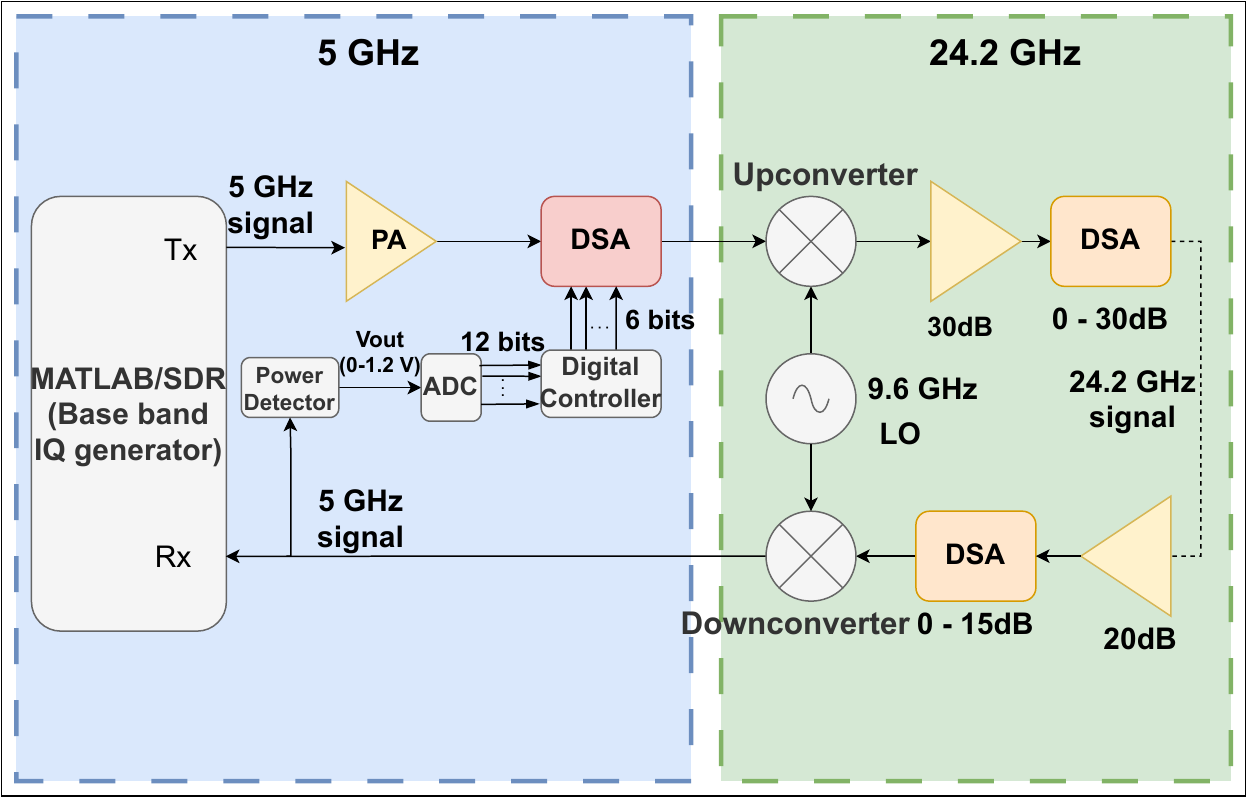}}
\caption{Block diagram of the 24 GHz mmWave transceiver with fuzzy-integral power control.}
\label{fig1}
\end{figure}

The transceiver architecture in Fig.~\ref{fig1} includes baseband modulation, up- and down-conversion stages, power amplification, and a closed-loop power control path, following approaches using fuzzy controllers for robust 5G performance~\cite{b2}. A software-defined radio (SDR) generates a 5~GHz IF 5G waveform, which a sub-harmonic mixer upconverts to 24.2~GHz using a 9.6~GHz local oscillator (LO). The signal passes through variable gain amplifiers (VGAs), digital step attenuators (DSAs), and a power amplifier (PA) for gain control.

\begin{figure*}[!t]
		\centerline{\includegraphics[width=\linewidth]{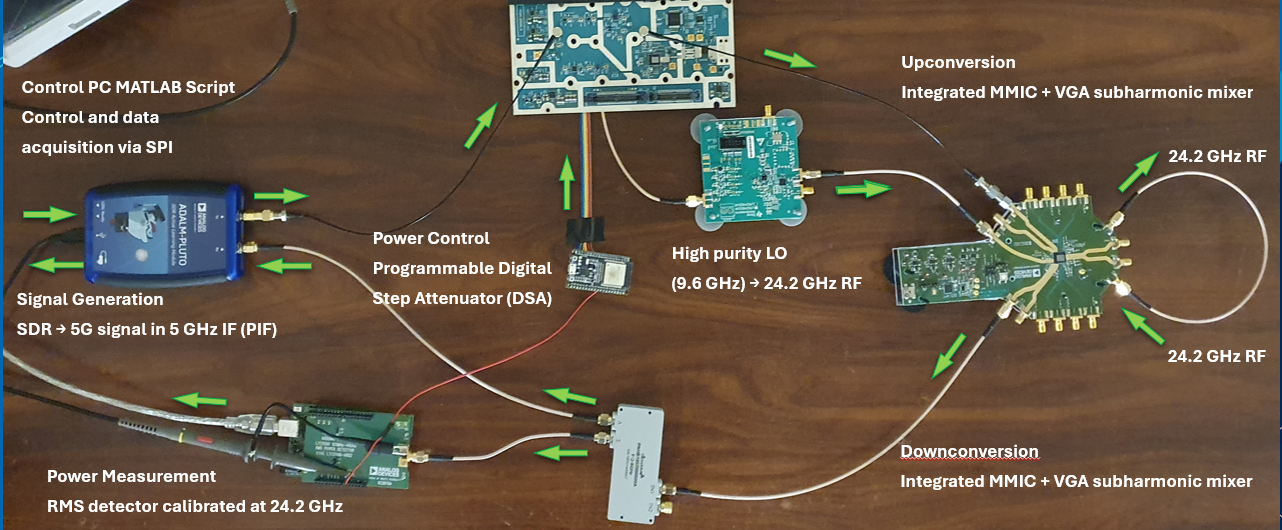}}
		\caption{Experimental setup showing the SDR, a power amplifier, a 6-bit variable attenuator, a local oscillator (LO) at 9.6 GHz, an up-converter mixer with an output at 24.2 GHz, a down-converter mixer to 5 GHz, a power detector, and reception in the SDR.}
		\label{figexp}
	\end{figure*}

The mixer output is modeled as the product of IF and effective LO signals, giving $f_{RF} = 2f_{LO} \pm f_{IF}$. Measurements were performed on the experimental setup implementing the block diagram of Fig.~\ref{fig1}, shown in Fig.~\ref{figexp}. In this setup, the SDR feeds the signal through a PA and a 6-bit variable attenuator (for impedance matching) before the mixer. With the 9.6~GHz LO, the signal is up- and down-converted, while a power detector monitors the RF output, which is finally received by the SDR.

Additionally, an RMS power detector monitors the RF output and generates a DC voltage proportional to the average power, following architectures similar to integrated mmWave power detection circuits \cite{b1}. This signal is fed into the analog-to-digital converter (ADC) of an ESP32 microcontroller, which executes the dynamic controller. The controller adjusts the DSA attenuation to maintain the target output power. The maintenance of constant power is effectively equivalent to maintaining a low RMS EVM; as shown in Section \ref{EVM_power}, at an average output power of $-30 dBm$, the EVM remains at an optimal value below $1.5\%.$

\section{RMS Power Detector}
The RMS power detector translates the mmWave signal into a DC voltage proportional to the average RF power, enabling precise monitoring for closed-loop control. The detector's response is characterized by the equation:
$ V_{\text{out}} = S \cdot (P_{\text{RF}} - P_0) $
where $ S $ is the slope in V/dB, $ P_{\text{RF}} $ is the input power in dBm, and $ P_0 $ is the intercept point in dBm.

\begin{figure}[htbp]
    \centerline{\includegraphics[width=0.48\textwidth]{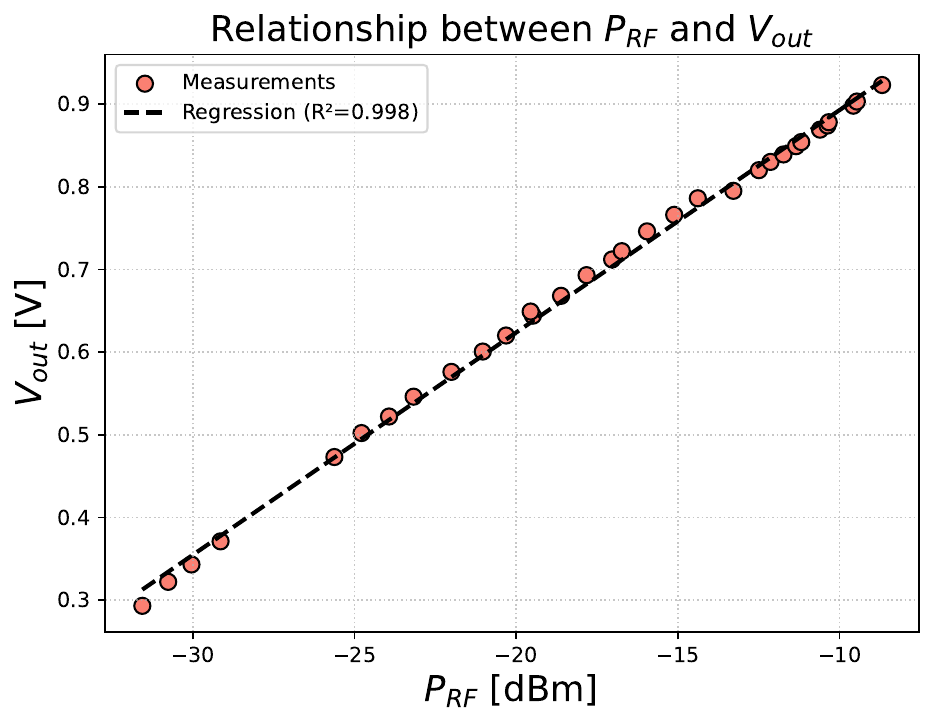}}
    \caption{Characterization of the RMS power detector showing output voltage versus input power with linear regression fit ($R^2=0.998$).}
    \label{detector_cal}
\end{figure}

Characterization was performed by sweeping the RF input power and measuring the output voltage. A linear regression fit yields the slope and intercept parameters. Fig. \ref{detector_cal} illustrates the measured data points and the regression line, confirming the log-linear behavior over the operational range.

\section{EVM RMS and Output Power relationship} \label{EVM_power}
To validate the relationship between output power stability and EVM preservation, the transceiver was first characterized in an open-loop configuration. The IF drive level from the SDR was swept while measuring the resulting RF output power at 24.2 GHz using the calibrated RMS detector. In parallel, the error vector magnitude (EVM) was evaluated after down-conversion and demodulation to quantify modulation fidelity across varying power levels.

\begin{figure}[htbp]
    \centerline{\includegraphics[width=0.48\textwidth]{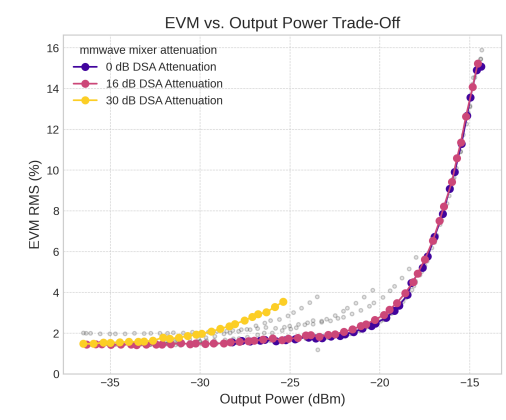}}
    \caption{Experimental power-linearity trade-off of the 24.2 GHz transmit chain. Higher gain settings achieve greater output power at the cost of accelerated non-linear EVM degradation.}
    \label{evm}
\end{figure}

The results confirm that the system maintains linear behavior up to approximately -30 dBm of output power, beyond which gain compression induces a rapid increase in EVM. This observation experimentally supports the control objective stated: maintaining constant output power inherently maintains low EVM. Consequently, -30 dBm is selected as the optimal operating point for closed-loop regulation, ensuring operation within the linear region where RMS EVM remains below $1.5\%$. Fig. \ref{evm} illustrates the power–EVM trade-off for different values of mixer attenuation. 

\section{Control Strategies Design}
This section details the design of the PID, pure integral (I), and fuzzy-integral (FI) controllers, each aiming to maintain PA output power at a target power by adjusting DSA attenuation.

\subsection{PID Controller}
The classical PID controller combines proportional, integral, and derivative actions for fast response, zero steady-state error, and oscillation damping. Its discrete form is:

\begin{equation}
    u[k] = K_p e[k] + K_i \sum_{i=0}^{k} e[i] T_s + K_d \frac{e[k] - e[k-1]}{T_s}
\end{equation}

where $ e[k] = P_{ref} - P_{RF}[k] $ is the error, $ T_s $ is the sampling time, and $ K_p, K_i, K_d $ are tuned gains. 

Gain tuning can follow various approaches; however, the RF chain's nonlinearities and the fast dynamics of amplifiers and attenuators make conventional methods largely impractical without specialized equipment. Therefore, iterative trial-and-error is used, assuming static RF gains (neglecting dynamics and considering only steady-state behavior), despite this differing from actual system dynamics.

\subsection{Pure Integral (I) Controller}
The pure integral controller, focuses on eliminating steady-state error without proportional or derivative terms:

\begin{equation}
    u[k] = u[k-1] + K_i e[k] T_s
\end{equation}

 Since the system is considered linear, stability is guaranteed, steady-state error is eliminated, and disturbances are rejected, all with a straightforward analytical development. In this case, the time constant — which can be used as the design parameter for the controller (instead of relying on trial-and-error tuning as in full PID control) — is given by:

\begin{equation}
    \tau = \frac{1}{\alpha \cdot K_i}
\end{equation}

where $\alpha$ denotes the overall RF-chain gain and $K_i$ is the integral constant. Note that this analysis assumes a fully linear and dynamics-free RF chain—neither strictly valid. The controller operates around an equilibrium.

\subsection{Fuzzy-Integral (FI) Controller}
The FI controller addresses system nonlinearities and uncertainties. Fuzzy logic mimics human decision-making using linguistic variables and membership functions, while the integral term eliminates steady-state error. The design follows principles of robust fuzzy logic schemes applied in wireless networks to handle signal variability and uncertain conditions \cite{b4,b3}.

Inputs are power error $ e = P_{\text{ref}} - P_{\text{RF}} $ and its change $ \Delta e = de/dt $. Output is incremental attenuation adjustment $ \Delta u $, integrated to yield total attenuation $ u = \int \Delta u \, dt $.

Triangular membership functions are used for seven linguistic terms: negative big (NB), negative medium (NM), negative small (NS), zero (Z), positive small (PS), positive medium (PM), positive big (PB).

The rule base in Table I employs Mamdani inference using \textit{min} for AND, \textit{max} for OR, and centroid defuzzification. Each cell defines the linguistic mapping from the input pair $(e, \Delta e)$ to the output $\Delta u$. For each power error and rate combination, the rule determines the incremental attenuation adjustment. The controller simultaneously evaluates all active rules with \textit{min}–\textit{max} operators, and centroid defuzzification yields the final crisp output $\Delta u$, ensuring smooth, adaptive control of the system’s nonlinear behavior.

\begin{table}[htbp]
\caption{Fuzzy Rule Base}
\centering
\renewcommand{\arraystretch}{1.3} 
\setlength{\arrayrulewidth}{0.8pt} 
\arrayrulecolor{black}

\begin{tabular}{|c|c|c|c|c|c|c|c|}
\hline\hline
\cellcolor{gray!20}\textbf{$e$} 
 & \multicolumn{7}{c|}{\cellcolor{gray!20}\textbf{$\Delta e$}} \\ \cline{2-8}
\cellcolor{gray!20} 
 & \cellcolor{gray!10}\textbf{NB} 
 & \cellcolor{gray!10}\textbf{NM} 
 & \cellcolor{gray!10}\textbf{NS} 
 & \cellcolor{gray!10}\textbf{Z}  
 & \cellcolor{gray!10}\textbf{PS} 
 & \cellcolor{gray!10}\textbf{PM} 
 & \cellcolor{gray!10}\textbf{PB} \\ \hline\hline
\cellcolor{gray!10}\textbf{NB} & NB & NB & NB & NB & NM & NS & Z  \\ \hline
\cellcolor{gray!10}\textbf{NM} & NB & NB & NB & NM & NS & Z  & PS \\ \hline
\cellcolor{gray!10}\textbf{NS} & NB & NB & NM & NS & Z  & PS & PM \\ \hline
\cellcolor{gray!10}\textbf{Z}  & NB & NM & NS & Z  & PS & PM & PB \\ \hline
\cellcolor{gray!10}\textbf{PS} & NM & NS & Z  & PS & PM & PB & PB \\ \hline
\cellcolor{gray!10}\textbf{PM} & NS & Z  & PS & PM & PB & PB & PB \\ \hline
\cellcolor{gray!10}\textbf{PB} & Z  & PS & PM & PB & PB & PB & PB \\ \hline\hline
\end{tabular}
\label{tab1}
\end{table}



\section{Experimental Results And Discussion}

The experimental setup shown in Fig. \ref{fig1} is used to test all the controllers. The open-loop characterization established the linear operating region up to approximately -30 dBm, where EVM begins to degrade due to PA compression. For closed-loop evaluation, the target output power was set to -30 dBm to ensure operation well within the linear regime, minimizing EVM and spectral regrowth.

\subsection{Disturbance Rejection Performance}

The disturbance rejection experiments emulate real-world operational perturbations, such as variations in load or environmental conditions, by abruptly changing the mmWave link attenuation. Two test scenarios were applied: (1) a step decrease in attenuation from 10~dB to 5~dB (equivalent to a gain increase), and (2) a step increase from 5~dB to 10~dB (gain reduction). These abrupt transitions are representative of the large-signal, nonlinear disturbances typically encountered in high-frequency RF front-ends.

\begin{figure*}[!t]
    \centering
    \includegraphics[width=\textwidth]{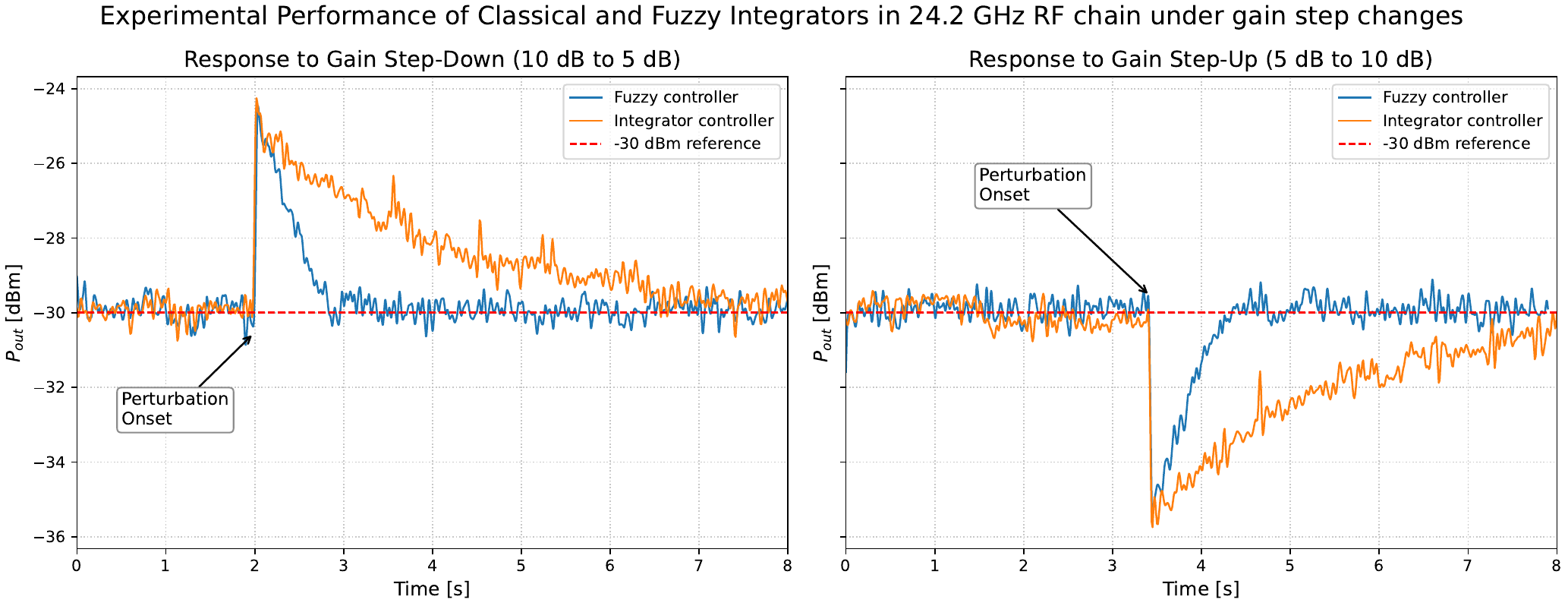}
    \caption{Temporal response of classical and fuzzy integrator controllers managing the output power of the 24.2 GHz RF system. The chain is subject to abrupt variations in system gain. The left subplot illustrates the transient response following a step gain reduction from 10 dB to 5 dB. Conversely, the right subplot displays the response to a step gain increase from 5 dB to 10 dB. Both control architectures utilize an identical integral gain ($K_i = 2$). In both scenarios, the fuzzy integrator demonstrates a significantly faster transient response compared to its conventional counterpart.}
    \label{fig5}
\end{figure*}


Fig.~\ref{fig5} compares the transient responses of the fuzzy–integral (FI) and pure integral (I) controllers under identical conditions and with the same integral gain. The FI controller settles in approximately $1~\text{s}$ with negligible overshoot and no oscillation, whereas the I controller requires nearly $5~\text{s}$ to reach steady state. This fivefold improvement highlights the FI controller’s ability to adapt its effective gain through fuzzy inference, accelerating convergence without loss of steady-state accuracy. The fuzzy rule base’s adaptive mapping mitigates nonlinearities and gain variations, enabling high responsiveness even when the plant departs from the linear region.

\begin{figure}[htbp]
    \centerline{\includegraphics[width=0.48\textwidth]{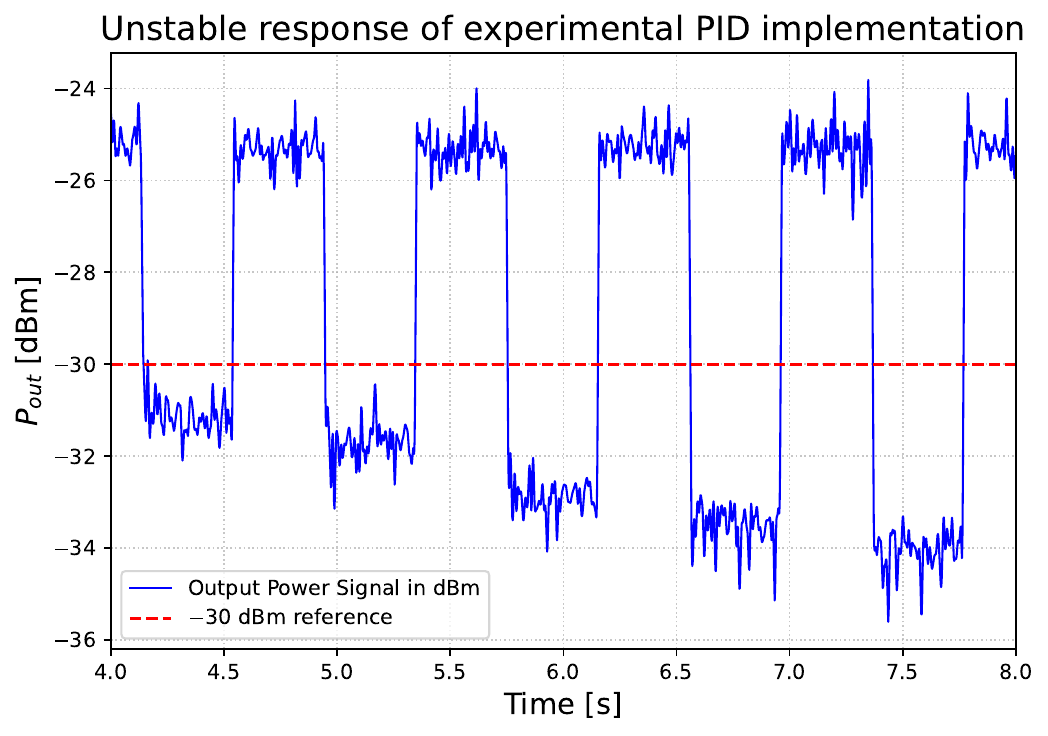}}
    \caption{Closed-loop output power exhibits a sustained limit cycle around the $-30\,\mathrm{dBm}$ reference when controlled by a conventional PID. Subunit variations in $K_P$ and $K_D$ destabilize the loop, consistent with the controller's local stability on nonlinear mmWave RF plants where PA compression and gain drift shift the effective loop gain and phase margin.
}
    \label{fig6}
\end{figure}

Conversely, the PID controller response is highly sensitive to the proportional ($K_p$) and derivative ($K_d$) gains, where small variations ($<1$ unit) cause oscillations or ringing, as shown in Fig.~\ref{fig6}, consistent with the local stability analysis in Section~\ref{theo_expl}.

The pure integral (I) controller is stable but slow. The FI controller adds adaptive nonlinear compensation, achieving faster settling and better disturbance rejection. Experiments confirm its extended stability region and consistent performance under nonlinear, time-varying conditions.

\subsection{Theoretical Analysis of PID and I Inefficiencies}\label{theo_expl}

The limitations of PID and pure integral (I) controllers can be explained using nonlinear stability theory. Both rely on a locally linearized plant model around a nominal operating point, enabling linear control design but restricting stability guarantees to that region. In practice, the mmWave transceiver exhibits strong nonlinearities—mainly from power amplifier (PA) compression and temperature-induced gain variation—so linear controllers may become unstable when the system deviates from its nominal operating point.

Formally, the closed-loop system can be represented as $\dot{x} = f(x)$ with equilibrium $x_0$. Local asymptotic stability is defined as:

\begin{equation}
\exists\, \delta > 0 : \|x(0) - x_0\| < \delta \Rightarrow \lim_{t \to \infty} x(t) = x_0,
\end{equation}

as established in standard nonlinear control theory \cite{khalil,slotine}. This implies that stability is only ensured within a small neighborhood of $x_0$; beyond that region ($\|x(0) - x_0\| > \delta$), no guarantee exists. Therefore, when a PID or I controller is tuned empirically based on a locally linear model, its stability can degrade rapidly as the plant dynamics shift due to nonlinear effects or parameter drifts.

For the I controller, $C(s) = K_I/s$, the loop gain $L(s) = K_I \alpha / s$ ensures zero steady-state error under small perturbations but introduces a phase lag of $\pi$ radians, which—combined with neglected high-frequency dynamics—may trigger oscillations if the system gain $\alpha$ varies. Similarly, the PID controller, $C(s) = K_P + K_I/s + K_D s$, can provide faster transients but is highly sensitive to unmodeled nonlinearities; excessive $K_P$ or $K_D$ reduces the phase margin, leading to sustained oscillations observed experimentally.

Consequently, while both controllers can stabilize the system locally, their robustness is fundamentally limited by the assumption of linearity. As operating conditions drift away from the equilibrium point, stability may no longer hold. This motivates the use of nonlinear strategies, such as the fuzzy–integral (FI) controller, which adaptively reshapes the control surface based on real-time inference, preserving stability and performance even under nonlinear or time-varying conditions.

\section{Conclusion}
This paper introduced a comparative study of PID, pure integral, and fuzzy-integral (FI) controllers for adaptive power management in a 24 GHz mmWave 5G transceiver. The FI approach, integrating fuzzy logic’s robustness with integral action, offers superior performance with quicker settling times and no overshoot, suggesting its advantage in complex RF environments.

These findings support reliable high-data-rate 5G communications with minimal EVM degradation and spectral regrowth. Limitations include computational overhead on resource-constrained devices, though the ESP32 implementation proves feasibility. Future work will explore integration with digital predistortion (DPD) techniques to enhance linearity at higher power levels.

\end{document}